\documentclass[doublecol]{epl2} 
\usepackage{graphicx}
\usepackage{amsmath}
\usepackage{amssymb}
\usepackage{ulem}
\usepackage[subnum]{cases}

\newcommand{\bs}[1]{{\boldsymbol{#1}}}

\newcommand{\bq}{\bs{q}}

\newcommand{\be}{\bs{\epsilon}}

\title{Casimir-Polder force fluctuations as spatial probes of dissipation in metals}
\shorttitle{Casimir-Polder force fluctuations as spatial probes of dissipation in metals} %Insert here a short version of the title if it exceeds 70 characters

\author{Nicolas Cherroret\footnote{Contact: cherroret@lkb.upmc.fr} \and Pierre-Philippe Cr\'epin \and Romain Gu\'erout, Astrid Lambrecht \and Serge Reynaud}
\shortauthor{N. Cherroret \etal}

\institute{                    
Laboratoire Kastler Brossel, UPMC-Sorbonne Universit\'es, CNRS, ENS-PSL Research University, Coll\`{e}ge de France, 4 Place Jussieu, 75005 Paris, France
}

\abstract{
We study the spatial fluctuations of the Casimir-Polder force experienced by an atom or a small sphere moved above a metallic plate at fixed separation distance. We demonstrate that unlike the mean force, the magnitude of these fluctuations crucially relies on the  relaxation of conduction electron in the metallic bulk, and even achieves values that differ by orders of magnitude depending on the amount of dissipation. We also discover that fluctuations suffer a spectacular decrease at large distances in the case of nonzero temperature.
}

\begin{document}

\maketitle

\section{Introduction}

The Casimir effect constitutes a paradigmatic example of dispersion force between neutral bodies, induced by quantum fluctuations of the electromagnetic field. Since their discovery \cite{Casimir48, CasimirP48}, Casimir and Casimir-Polder forces have had a large impact in the fields of physics, chemistry, biology and nanotechnology \cite{Woods16,Parsegian2006, Milton2011}.
The basics of the Casimir effect has recently attracted renewed interest, as a result of a large amount of experimental work allowing for precision measurements, and of the observation of disagreements between the results of these experiments and theoretical predictions \cite{Lambrecht2011}. 
%\cite{LamoreauxPRL1997, MohideenPRL1998,
%EderthPRA2000,ChanPRL2001,ChenPRL2002,DeccaPRL2003,Decca07,vanZwolAPL2008,
%ChanPRL2008,JourdanEPL2009,Munday09,MasudaPRL2009,deManPRL2009,ChanPRL2010,
%TorricelliEPL2011, SushkovNatPh2011,TangPRL2012,DeccaNatComm2013}.

Most precise measurements of the Casimir force are performed between large spheres and metallic plates separated in distances ranging from a fraction of micrometer to a few micrometers. The force is dominated by zero-point quantum fluctuations of the electromagnetic field at separations much smaller than the thermal wavelength $\lambda_T=\hbar c/(k_B T)$ ($7.6\mu$m at room temperature), whereas thermal fluctuations also contribute at larger separations. In both cases, the magnitude of the force depends on the reflection properties of the sphere and plate, which themselves depend on the complex dielectric function of the materials. In metals, the low-frequency limit of the latter is  controlled by the conductivity, that is in practice by the Drude description of electron scattering from the metal impurities.
The most precise measurements performed at submicrometric separations appear to be in good agreement with the so-called plasma model, that is the Drude model with dissipation discarded \cite{Decca07, Chang12, Banishev13, Bimonte16}.
In contrast, experiments performed at larger distances of a few micrometers (i.e. at separations approaching $\lambda_T$) obtain results in good agreement with the dissipative Drude model, after a large contribution of electrostatic effets is substracted \cite{SushkovNatPh2011, TangPRL2012}.   To the best of our knowledge, this intriguing contradiction on the role of dissipation in Casimir experiments  has not yet been solved, though a number of potential explanations has already been investigated \cite{Reynaud2013}.

\begin{figure}
\includegraphics[width=1\linewidth]{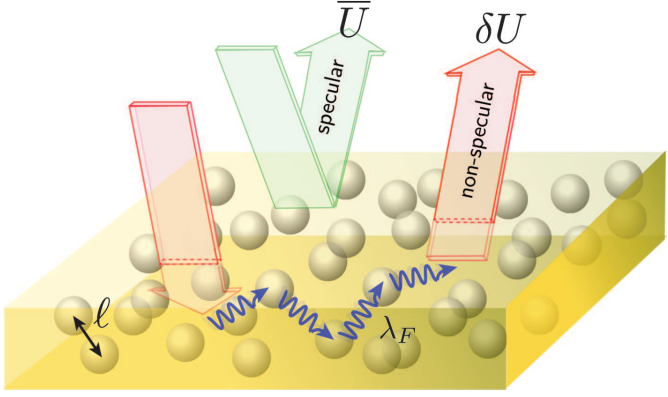}
\caption{
\label{illustration}
(color online) The Casimir potential $U=\overline{U}+\delta U$ between an arbitrary object and a metallic plate containing impurities has two contributions. The first is the (main) specular part, $\overline{U}$, controlled by the reflection properties of the flat surface of the metal. The second is a non-specular part, $\delta U$, and originates from the spatial heterogeneities (impurities) in the metallic bulk, from which conduction electrons (of Fermi wavelength $\lambda_F$) are scattered (with a mean free path $\ell$).
}
\end{figure}

In this Letter we propose to study a new effect, related to the fluctuations of the Casimir-Polder (CP) force, which could help to improve our understanding of the role of dissipation in Casimir physics. Considering the geometry of a small sphere (or an atom)  located at a distance $z$ above a plane metallic plate, we study the \textit{non-specular} contribution to the CP interaction potential $U$ (from which the force $F=-\partial U/\partial z$ is derived), which is inherently connected with electron relaxation within the metal. In practice, this contribution manifests itself as a spatial fluctuation $\delta U$ of the potential around a mean value $\overline{U}$ as the sphere is moved above the plate at fixed separation. Alternatively, by virtue of the ergodicity principle, $\delta U$ can also be seen as a fluctuation from sample to sample of the Casimir interaction potential between a sphere and macroscopically identical --but microscopically different-- metallic plates. Physically, the mean CP potential $\overline{U}$ is mediated by the field fluctuations specularly reflected from the surface of the metal. The space-varying fluctuation $\delta U$, on the other hand, stems from the part of the radiation that is re-emitted \textit{non specularly}, as illustrated in Fig. \ref{illustration}, after having penetrated the metal and interacted with electrons scattered in the disordered metallic bulk.
Due to this mechanism, the very existence of the  fluctuating part of the Casimir force relies on the presence of impurities in the metallic plate, and thus constitutes a natural probe of dissipation.

In the following, we provide for the first time a detailed description of these spatial fluctuations for a metallic plate, which turn out to be richer than the fluctuations arising with  dielectric disordered materials \cite{Dean09, Cherroret15, Cherroret15b}. We discover that unlike the mean CP potential, at $T=0$ and large distances the variation of these fluctuations with the sphere-plate separation  distance significantly depends on the amount of dissipation in the metallic bulk. Furthermore, by analyzing $\delta U$ at finite temperature, we come up with the surprising result that fluctuations decay exponentially with the sphere-plate separation, in strong contrast with the behavior of mean Casimir potentials that decay algebraically.

\section{Scattering approach}

Let us consider the CP interaction potential $U$ between a small resonant dielectric sphere of polarizability $\alpha(\omega)=\alpha(0)\omega_0^2/(\omega_0^2-\omega^2)$, located at a distance $z$ from a semi-infinite metallic plate. The choice of a dielectric sphere is made for simplicity, and does not involve any loss of generality. All the results presented in the Letter apply to a two-level atom or a small metallic sphere as well, the role of the resonant frequency $\omega_0$ being then played by the plasma frequency of the sphere.
At zero temperature and in the dipolar approximation for the sphere, $U$ is given by \cite{Emig07, Messina09}
\begin{eqnarray}
U&=&\Im
\left[-\frac{\hbar}{\epsilon_0 c^2}\int_0^\infty\frac{d\omega}{2\pi}i\omega^2\alpha(\omega)
\int\frac{d^2\bq_a}{(2\pi)^2}
\frac{d^2\bq_b}{(2\pi)^2}
\frac{e^{i (k_a^z+k_b^z)z}}{2 k_a^z}
\right.\nonumber\\
&&\times \sum_{a,b}r_{ab}(\omega)
\, \be_a^+(\bq_a)\cdot\be_b^-(\bq_b)\bigg],
\label{EqU}
\end{eqnarray}
where $r_{ab}(\omega)$ is the reflection coefficient of the plate, describing scattering of an incoming mode with transverse wave vector $\bq_a$ and polarization vector $\be_a^+(\bq_a)$ into an outgoing mode $\{\bq_b,\ \be_b^-(\bq_b)\}$, at frequency $\omega$. $a,b\in\left\{\text{TE},\text{TM}\right\}$ are polarization indices, and $(k^z_a)^2=(\omega/c)^2-\bq_a^2$. 
Within an effective-medium description, the metal is purely homogeneous and electromagnetic fields are specularly reflected from the surface, with reflection amplitudes $\overline{r}_{ab}(\omega)=(2\pi)^2\delta(\bq_1-\bq_b)\delta_{ab} r_a(\omega)$, where $r_a(\omega)$ are the Fresnel coefficients of the vacuum-metal interface. These coefficients depend on the (Drude) complex permittivity of the metal, $\epsilon(\omega)=1-\omega_p^2/[i\omega(\gamma-i\omega)]$, where $\omega_p$ is the plasma frequency and $\gamma$ the electron relaxation rate in the metal. At large separations $z\gg\lambda_p=2\pi c/\omega_p$, this leads to the known result $\overline{U}=-3\hbar c \alpha(0)/(32\pi^2\epsilon_0z^4)$ \cite{Messina09}.
%\begin{equation}
%\overline{U}=-\frac{3\hbar c \alpha(0)}{32\pi^2\epsilon_0z^4}.
%\label{Ubar_largez}
%\end{equation}
Note that this relation is independent of $\gamma$, and in particular holds when $\gamma\to0$ (plasma model).

\begin{figure}[h]
\includegraphics[width=1\linewidth]{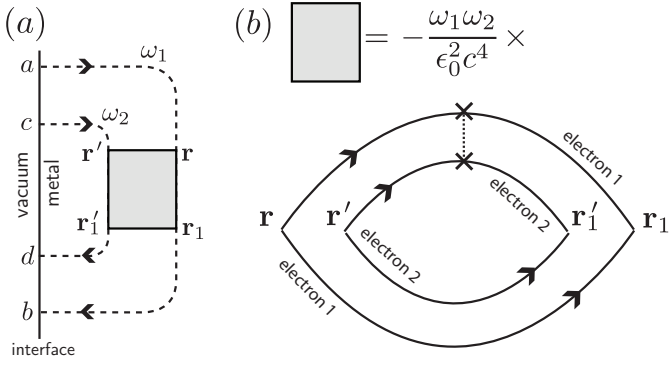}
\caption{
\label{diagrams}
(a) Diagram representing the correlator $\overline{\delta r_{ab}(\omega_1)\delta r_{cd}(\omega_2)}$. Dashed lines symbolize propagation of the electromagnetic field inside the metal. $a$ and $c$ are the incoming modes, and $b$ and $d$ the outgoing modes. The field in the outer (inner) branch accelerates an electron at point $\textbf{r}$ ($\textbf{r}'$), which later re-emits a radiation at point $\textbf{r}_1$ ($\textbf{r}_1'$). (b) The conductivity correlator $\overline{\delta \sigma_{ab}(\textbf{r},\textbf{r}_1,\omega_1)\delta \sigma_{cd}(\textbf{r}',\textbf{r}_1',\omega_2)}$ encodes the correlation of the two electron scattering trajectories.
For a good metal $(k_F\ell \gg 1)$, the leading-order contribution to this correlator involves only one scattering event on an impurity of the metal (cross symbol). }
\end{figure}

$\overline{U}$ is not the only contribution to $U$. Indeed, the presence of impurities in the metal makes  the CP potential $U=\overline{U}+\delta U$ fluctuate spatially around its mean value $\overline{U}$. The fluctuating contribution $\delta U$ stems from electromagnetic fields that enter the metallic bulk  and are reflected through their interaction with conduction electrons scattered from impurities. We describe $\delta U$ by adding a non-specular contribution $\delta r_{ab}$ to the reflection coefficient, $r_{ab}=\overline{r}_{ab}+\delta r_{ab}$, which we calculate by making use of a statistical approach where the metallic plate is taken from a random ensemble of plates with different microscopic configurations of the impurity positions. The magnitude of potential fluctuations is then given by the variance $\overline{\delta^2U}\equiv \overline{U^2}-\overline{U}^2$, where the overbar denotes averaging over the random ensemble. 
To evaluate $\overline{\delta^2U}$, we square Eq. (\ref{EqU}), substract $\overline{U}^2$ and apply the configuration average. We obtain:
\begin{equation}
\begin{split}
\label{deltaUdef}
\overline{\delta U^2}(z)=\dfrac{-\hbar^2}{\epsilon_0^2c^4}
\text{Re}\bigg[
\int_0^\infty\dfrac{d\omega_1}{2\pi}\dfrac{d\omega_2}{2\pi}
\!\!\prod_{i=a,b,c,d}\!\int\dfrac{d ^2\bq_i}{(2\pi)^2}
\\
\sum_{a, b, c, d} \omega_1^2\omega_2^2
\alpha(\omega_1)\alpha(\omega_2)
\overline{\delta r_{ab}(\omega_1)\delta r_{cd}(\omega_2)}
\\
\times\dfrac{e^{i(k_a^z+k_b^z+k_c^{z}+k_d^{z})z}}{4k_{a}^zk_{c}^{z}} 
\be_a^+(\bq_a)\!\cdot\!\be_b^-(\bq_b)
\,\be_c^+(\bq_c)\!\cdot\!\be_d^-(\bq_d)
\bigg].
\end{split}
\end{equation}
Eq. (\ref{deltaUdef}) involves the correlator $\overline{\delta r_{ab}(\omega_1)\delta r_{cd}(\omega_2)}$ of reflection coefficients \cite{Cherroret15}, whose diagrammatic representation is shown in Fig. \ref{diagrams}a.
The outer branch of the diagram describes the following physical mechanism: an electromagnetic field fluctuation of frequency $\omega_1$ penetrates the metal, then accelerates an electron at some point $\textbf{r}$. This electron propagates through the metal up to a final point $\textbf{r}_1$ 
where it re-emits a radiation that eventually leaves the metal. The inner branch of the diagram describes a similar process at frequency $\omega_2$.
The correlator $\overline{\delta r_{ab}(\omega_1)\delta r_{cd}(\omega_2)}$ stems from the fact that the two electron scattering trajectories from $\textbf{r}$ to $\textbf{r}_1$ and from $\textbf{r}'$ to $\textbf{r}_1'$ can share one or several metal impurities. These trajectories are described by the conductivity fluctuations $\delta \sigma_{ab}(\textbf{r}, \textbf{r}_1,\omega_1)$ and $\delta \sigma_{cd}(\textbf{r}', \textbf{r}_1',\omega_2)$, so that their correlation is in turn encoded in the  correlator $\overline{\delta \sigma_{ab}(\textbf{r},\textbf{r}_1,\omega_1)\delta \sigma_{cd}(\textbf{r}',\textbf{r}_1',\omega_2)}$. The diagram in Fig. \ref{diagrams}a  explicitly reads:
\begin{equation}
\begin{split}
\overline{\delta r_{ab}(\omega_1)\delta r_{cd}(\omega_2)}=
4 k_a^z k_c^z
 \int  d^3\textbf{r}\, d^3\textbf{r}_1d^3\textbf{r}'d^3\textbf{r}_1'\\
\overline{G}_{\!a}(\bq_a,\textbf{r})
 \overline{G}_{b}(\bq_b,\textbf{r}_1)\overline{G}_{c}(\bq_c,\textbf{r}')\overline{G}_{d}(\bq_d,\textbf{r}_1')\\
\times
-\frac{\omega_1\omega_2}{\epsilon_0^2c^4}\overline{\delta \sigma_{ab}(\textbf{r},\textbf{r}_1,\omega_1)\delta \sigma_{cd}(\textbf{r}',\textbf{r}_1',\omega_2)}.
\label{dr_dr_def}
\end{split}
\end{equation}
In this expression, $\overline{G}_a(\bq_a,\textbf{r})=t_a^\text{vm}(1+r_a^\text{mm})\times 1/(2i \tilde{k}_a^z)e^{i\bq_a\cdot \textbf{r}_\perp+i \tilde{k}_a^z z}\be_a^+$ is the electromagnetic Green function that describes propagation of a radiation with polarization $\be_a^+$ and transverse (resp. longitudinal) wave vector $\bq_a$ ($\tilde{k}_a^z$) from the vacuum-metal interface to the point $\textbf{r}=(\textbf{r}_\perp, z)$ in the metal (with similar definitions for $\overline{G}_b$, $\overline{G}_c$ and $\overline{G}_d$). 
The longitudinal wave number $\tilde{k}_a^z$ is the one in the metal, defined as $(\tilde{k}_a^z)^2=(\omega_1/c)^2\epsilon(\omega_1)-\bq_a^2$ (and similarly for $\tilde{k}_b^z,\ \tilde{k}_c^z\ \tilde{k}_d^z$).
$t_a^\text{vm}$ is the Fresnel transmission coefficient from the vacuum to the metal, and $r_a^\text{mm}$ is the Fresnel refection coefficient from the metal to the metal (from now on, we omit the frequency dependence of these coefficients to lighten the notations). Physically, $t_a^\text{vm}$ accounts for the finite probability for a field fluctuation to penetrate the metal, while $r_a^\text{mm}$ accounts for the possibility for a field fluctuation inside the metal to be internally reflected from the surface. 
To evaluate  the conductivity correlator in Eq. (\ref{dr_dr_def}), we assume that the material is weakly disordered, i.e. that $k_F\ell\gg 1$, which is an excellent approximation for usual metals. The main contribution of the conductivity correlator to $\delta U$ is then due to electron trajectories correlated via a \textit{single} impurity \cite{Cherroret15}, as diagramatically shown in Fig. \ref{diagrams}b. Evaluation of this diagram leads to (see \cite{Kane88} for a more general discussion of conductivity correlations in metals):
\begin{eqnarray}
&&\hspace{-1cm}
\overline{\delta \sigma_{ab}(\textbf{r},\textbf{r}_1,\omega_1)\delta \sigma_{cd}(\textbf{r}',\textbf{r}_1',\omega_2)}= 
\dfrac{\delta_{ab}\delta_{cd}}{(1-i\omega_1/\gamma)^2(1-i\omega_2/\gamma)^2}\nonumber\\ 
&&\times\frac{\lambda_F^2\ell}{2\pi}\sigma_0^2
\delta(\textbf{r}-\textbf{r}')\delta(\textbf{r}_1-\textbf{r}_1')\delta(\textbf{r}-\textbf{r}_1),
\label{ds_ds}
\end{eqnarray}
where $\ell$ is the electron mean free path ($\ell=v_F/\gamma$ with $v_F$ the Fermi velocity), $\lambda_F$ the Fermi wavelength and $\sigma_0=\epsilon_0\omega_p^2/\gamma$ is the Drude conductivity. To derive Eq. (\ref{ds_ds}), we have neglected the finite range and anisotropic structure of the conductivity correlator \cite{Kane88}. Taking into account this structure is not necessary here, as it would eventually give rise to relative corrections to $\overline{\delta^2 U}$ smaller by a factor $\sim 1/(k_F\ell)\ll 1$.
By reporting Eq. (\ref{ds_ds}) into (\ref{dr_dr_def}), we obtain
\begin{equation}
\begin{split}
\overline{\delta r_{ab}(\omega_1)\delta r_{cd}(\omega_2)}=
\frac{\pi \lambda_F^2\ell\omega_p^4}{2 c^4\gamma^2}
\frac{\omega_1\omega_2}{(1-i\omega_1/\gamma)^2(1-i\omega_2/\gamma)^2}\\
\times\frac{-ik_a^zk_c^z (\be_a^+\!\cdot\!\be_b^-)(\be_c^+\!\cdot\!\be_d^-)}{\tilde{k}_a^z\tilde{k}_c^z\tilde{k}_b^z\tilde{k}_d^z(\tilde{k}_a^z +\tilde{k}_b^z+\tilde{k}_c^{z}+\tilde{k}_d^{z})}
\delta(\bq_a\!-\!\bq_b\!-\!\bq_c\!+\!\bq_d)
\\\times t_a^\text{vm}t_c^\text{vm}
t_b^\text{mv}t_d^\text{mv}
\prod_{i=a,b,c,d}(1+r_i^\text{mm}),
\label{dr_dr}
\end{split}
\end{equation}
where $t_b^\text{mv}$, $t_d^\text{mv}$ are the Fresnel transmission coefficients from the metal to the vacuum. The Dirac delta function that appears in Eq. (\ref{dr_dr}) signals momentum conservation of the total scattering process.
%The reflection coefficients $r_i^\text{me}$, on the other hand, accounts for the possibility for the field inside the metal to be internally reflected from the surface. 

\section{Results}

\begin{figure}[h]
\includegraphics[width=1\linewidth]{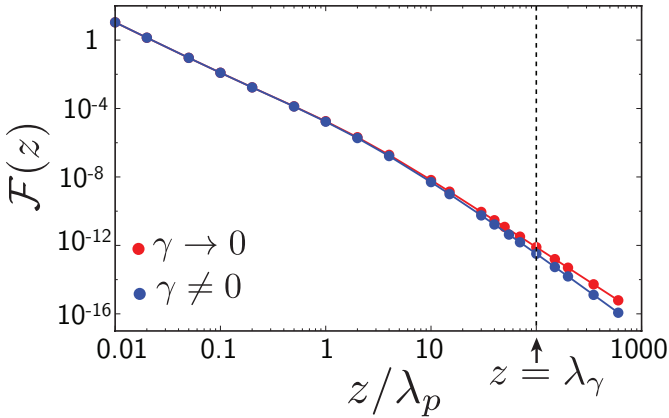}
\caption{
\label{Drude_vs_Plasma}
(color online) Function $\mathcal{F}(z)$ versus $z/\lambda_p$, at $T=0$ and for $\lambda_0=\lambda_p$ (blue dots). We have set $\lambda_\gamma=2\pi c/\gamma=10^2\lambda_0$. Red dots show the same function in the limit $\gamma\to 0$. Lines joining the points are guides to the eye.}
\end{figure}

By computing $\overline{\delta^2U}$ using Eqs. (\ref{deltaUdef}) and (\ref{dr_dr}), we obtain the final expression
\begin{equation}
\frac{\overline{\delta^2U}}{\overline{U}^2}=
\frac{(2\pi\lambda_F)^2\ell}{\lambda_\gamma^2\lambda_p}\, \mathcal{F}(z),
\label{deltaU_U}
\end{equation}
where $\lambda_\gamma=2\pi c/\gamma$.
The prefactor $(2\pi\lambda_F)^2\ell/(\lambda_\gamma^2\lambda_p)$ describes the interaction of the electromagnetic field with conduction electrons in the metal, and quantifies the strength of relative fluctuations. Its physical interpretation will be elucidated later. At this stage, let us mention that the function $\mathcal{F}(z)$ remains finite when $\gamma\to 0$ (see below). Therefore, since $(2\pi\lambda_F)^2\ell/(\lambda_\gamma^2\lambda_p) \propto \gamma$, the fluctuating part of the CP potential vanishes when $\gamma\to 0$. This is required since the spatial fluctuations of $U$ find their origin in the metal heterogeneities, encoded in $\gamma$.  
We note that this important property was not fulfilled in a recent work by Allocca et al. \cite{Allocca15}, who used a diffusion approximation to describe electronic motion in a metal. Such an approximation is inadequate in the context of the Casimir effect because it underestimates the weight of short electron scattering trajectories, which turn out to be the dominant ones \cite{Cherroret15}. The function $\mathcal{F}(z)$ is displayed in Fig. \ref{Drude_vs_Plasma} (blue dots). For comparison, we also show the same function in the limit where $\gamma\to 0$ (red dots), calculated using the plasma limit for the mean permittivity $\epsilon(\omega)$.
For simplicity we here set $\lambda_0=\lambda_p$, so that only two characteristic length scales remain in the problem, the plasma wavelength $\lambda_p$ and the relaxation wavelength $\lambda_\gamma=2\pi c/\gamma$. 
Overall, $\mathcal{F}(z)$ is a decaying function of $z$, which  confirms the intuition that the sphere tends to average out the metal heterogeneities at large separations.
At small separations $z\ll\lambda_p$, we find $\mathcal{F}(z)\propto (\lambda_p/z)^3$. This characteristic scaling is not surprising, as it is reminiscent to what has been found recently for spatial fluctuations of CP forces above dielectric disordered plates \cite{Dean09, Cherroret15}. Indeed, small separations are described by large frequencies where the electromagnetic field penetrates easily the metal, which thus behaves similarly to a dielectric material. 
\begin{figure}[h]
\includegraphics[width=1\linewidth]{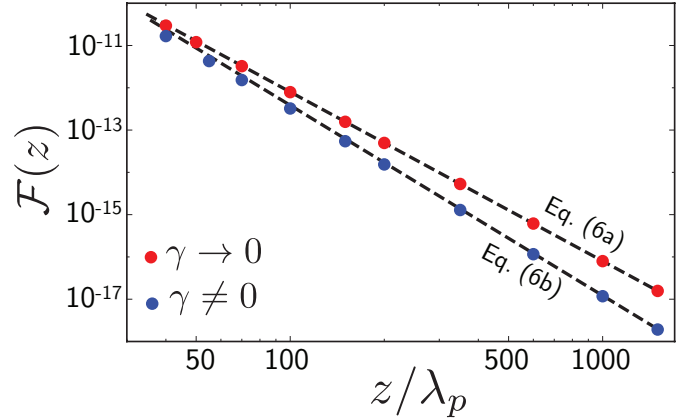}
\caption{
\label{Drude_vs_Plasma_zoom}
(color online) Zoom of figure \ref{Drude_vs_Plasma} in the range $z\gg\lambda_p$.
The dashed lines are the asymptotes (\ref{Fz_asymptoticsa}) and (\ref{Fz_asymptoticsb}).
}
\end{figure}

A quick look at the two curves in Fig. \ref{Drude_vs_Plasma} also indicates that the function $\mathcal{F}(z)$ does not depend much on $\gamma$ at small separations $z\ll\lambda_p$. This can be qualitatively understood from the coincidence of both the plasma and the Drude models for the mean permittivity at large frequencies: $\epsilon(\omega)\to1-\omega_p^2/\omega^2$ whatever $\gamma$. 
In strong contrast, a very interesting behavior shows up at large separations where the $z$ variation of $\mathcal{F}(z)$ starts to qualitatively depend on $\gamma$. This is well visible in Fig. \ref{Drude_vs_Plasma_zoom}, which focuses on the range $z\gg\lambda_p$. 
This observation is also confirmed by an asymptotic analysis of $\mathcal{F}(z)$ at large separations, which yields:
\begin{numcases}{\mathcal{F}(z)=}
c_1\left(\dfrac{\lambda_p}{z}\right)^4 & $\lambda_p\ll z\ll\lambda_\gamma$\label{Fz_asymptoticsa}\\
c_2\left(\dfrac{\lambda_p}{\lambda_\gamma}\right)^4\left(\dfrac{\lambda_\gamma}{z}\right)^{9/2}& $z\gg\lambda_\gamma$
\label{Fz_asymptoticsb}
\end{numcases}
where $c_1\simeq 8.0\times10^{-5}$ and $c_2\simeq 3.9\times10^{-5}$ are numerical constants. These asymptotic limits are shown in Fig. \ref{Drude_vs_Plasma_zoom} as dashed lines, and describe very well the exact numerical results. 
Eqs. (\ref{Fz_asymptoticsa}-b) constitute an important result of the Letter. They indicate that at large separations, the variance of the CP potential has a different scaling with $z$ depending on $\gamma$. In other words, the fluctuations
achieve values that differ by  orders of magnitude  depending on the amount of dissipation in the metal.
This behavior can be traced back to the low-frequency asymptotics of  the mean permittivity $\epsilon(\omega)=1-\omega_p^2/[i\omega(\gamma-i\omega)]$, which is crucially affected by $\gamma$. Eq. (\ref{dr_dr}) depends on $\epsilon(\omega)$ through the wave numbers and Fresnel coefficients. At low frequencies, while $t^\text{vm}$ and $1/\tilde{k}^z\propto 1/\omega^{1/2}$ when $\gamma\ne 0$, $t^\text{vm}$ and $1/\tilde{k}^z\propto \omega^0$ in the limit $\gamma\to 0$.   
This shows again that the spatial fluctuations of Casimir-Polder forces could be used as an efficient probe to unambiguously assess the effect of dissipation in the Casimir effect in metals. 

%c_1\dfrac{\lambda_p}{z}^4& z\gg\lambda_p\ \text{(plasma)}\
%c_2\left(\dfrac{\lambda_p}{\lambda_\gamma}\right)^4\left(\dfrac{\lambda_\gamma}{z}\right)^{1/2}& z\gg\lambda_\gamma\ \text{(Drude)}

So far we have discussed only zero temperature. The effect of finite temperatures can be simply accounted for by  replacing the frequency integral in Eq. (\ref{EqU}) by a discrete sum over Matsubara frequencies. This modifies the mean Casimir interaction potential at large distances%$z\gtrsim \lambda_T=\hbar c/(k_BT)$, the thermal wavelength
, according to \cite{Ellingsen09}
\begin{equation}
\overline{U}=-\frac{1}{16\pi}\frac{\hbar c \alpha(0)}{\epsilon_0\lambda_T z^3}.
\label{Ubar_FiniteT}
\end{equation}
The calculation of the variance $\overline{\delta^2U}$ at $T\ne 0$ follows the same lines as at $T=0$, except that it involves a \textit{double} sum over Matsubara frequencies $2\pi k_BT n/\hbar $ and $2\pi k_BT m/\hbar $, where $n,m$ are integers running from 0 to $\infty$. At finite temperature, we find that the general form (\ref{deltaU_U}) still holds, with the function $\mathcal{F}(z)$ now modified at separations $z\gtrsim\lambda_T$ as compared to the results of Fig. \ref{Drude_vs_Plasma}.
\begin{figure}[h]
\includegraphics[width=1.0\linewidth]{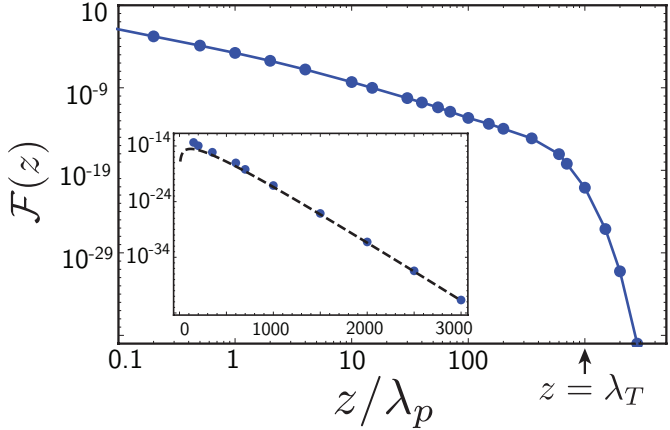}
\caption{
\label{RFluctuations_T}
(color online) Main panel: function $\mathcal{F}(z)$ at finite temperature, for $\lambda_0=\lambda_p$, $\lambda_\gamma=2\pi c/\gamma=10^2\lambda_0$ and  $\lambda_T=10^3\lambda_0$ (blue dots).
The fluctuations collapse exponentially at $z\gg\lambda_T$. Inset:
same curve in log scale, together with the asymptotic law (\ref{Asymptotics_T}).}
\end{figure}
$\mathcal{F}(z)$ is shown in the main panel of Fig. \ref{RFluctuations_T} at $T\ne0$ (blue dots). For definiteness we set  $\lambda_T=10^3\lambda_0>\lambda_\gamma=10^2\lambda_0$. Points such that $z\ll\lambda_T$ are identical to those in Fig. \ref{Drude_vs_Plasma}. At large separations however, temperature gives rise to a collapse of spatial fluctuations. A plot of $\mathcal{F}(z)$ in log scale (inset of Fig. \ref{RFluctuations_T}) suggests that this decay is close to \textit{exponential}.
Finite temperatures thus turn out to average out the spatial fluctuations of the Casimir force. To better understand this intriguing result, we have also investigated the regime $z\gg\lambda_T$ analytically. As is well known, for the mean potential $\overline{U}$ this limit is controlled by the zeroth-order Matsubara frequency term in the sum, which turns out to be nonzero and reduces to Eq. (\ref{Ubar_FiniteT}). The situation is very different for the variance $\overline{\delta^2 U}$, for which we find that the contributions involving zero-order  Matsubara frequencies ($n=0$ or $m=0$) identically \textit{vanish}. One can already guess this result from the expression of the correlator of reflection coefficients, Eq. (\ref{dr_dr}), which falls to zero when $\omega_1$ or $\omega_2\to 0$. 
Because the term ($n=0$ or $m=0$) is identically zero, the large-separation asymptotics of $\overline{\delta^2 U}$ is naturally governed by the first-order contribution $(n,m)=(1,1)$. This term can be explicitly calculated by means of a saddle-point approximation, leading to
\begin{equation}
\mathcal{F}(z)\simeq
c_3
\left(\dfrac{\lambda_p}{\lambda_T}\right)^4\left(1+\frac{\lambda_T}{\lambda_\gamma}\right)^{-1/2}\left(\frac{z}{\lambda_T}\right)^3e^{-8\pi z/\lambda_T},
\label{Asymptotics_T}
\end{equation}
where $c_3\simeq115.7$. This asymptote is shown in the inset of Fig. \ref{RFluctuations_T} as a dashed line, and matches well the exact numerical calculation. Note that the exponential decay of fluctuations is controlled by the thermal wavelength, the dissipation $\gamma$ appearing only  through the prefactor $(1+\lambda_T/\lambda\gamma)^{-1/2}$.
The result (\ref{Asymptotics_T}) is remarkably different from the algebraic decay of $\overline{U}$, Eq. (\ref{Ubar_FiniteT}). At a very qualitative level, it can be understood by the argument that in the strict limit of zero frequency, the electromagnetic field (of infinite wavelength) cannot resolve the spatial heterogeneities (impurities) of the metal. The contribution of zeroth-order Matsubara frequency to the fluctuation $\overline{\delta^2 U}$ must therefore be zero. We expect this argument to be universal, independent of the details of the material like the type of impurities or the value of the mean free path. It does of course not hold for the mean Casimir potential $\overline{U}$, which finds its origin in the reflection of the electromagnetic field from the purely \textit{homogeneous} surface of the metal, so that even a field of infinite wavelength can contribute to $\overline{U}$.

Having discussed how fluctuations behave as a function of the separation distance, let us now comment on the physical interpretation of the prefactor in Eq. (\ref{deltaU_U}), which controls their magnitude. Over a time span $t$, one can associate to an electron  trajectory in the metal an effective, classical tube of length $v_Ft$ and cross-section $\lambda_F^2$. This tube has a volume $V_e=v_F t\lambda_F^2$. Since a finite variance $\overline{\delta^2 U}$ arises due to correlations between electron scattering trajectories (see Fig. \ref{diagrams}), we have to estimate the probability for a crossing between to such tubes to take place. This probability is given by the ratio of $V_e$ to the effective volume $V$ of the metal accessible to the electromagnetic field. Over the same time span $t$, the field can transversally propagate over a surface $(ct)^2$ and it can penetrate the metal up to a distance $\sim\lambda_p$ (the typical skin depth), giving $V=\lambda_p(ct)^2$. 
The probability of crossing after a time $t$ is thus $(v_F t \lambda_F^2)/[\lambda_p(ct)^2 ]$. If we finally note  that the typical time scale for electron relaxation is $t=\gamma^{-1}$, that $v_F/\gamma\equiv\ell$ and $c/\gamma=\lambda_\gamma/(2\pi)$, we recover the prefactor in Eq. (\ref{deltaU_U}).

\section{Conclusion} 

In conclusion, we have described for the first time the spatial fluctuations of CP forces above metals. As these fluctuations are triggered by electron scattering, their observation would constitute a natural probe of dissipation in the Casimir effect. Furthermore, we have shown that their dependence on the sphere-plate separation depends significantly on the low-frequency description of the metal permittivity. For gold, a free electron density $n=6 \times 10^{28} / \text{m}^3$~\cite{AshcroftMermin} and an elastic mean free path $\ell = 37.7$ nm~\cite{Gall2016} lead to $[(2\pi\lambda_F)^2\ell/(\lambda_\gamma^2\lambda_p)]^{1/2}\sim 3.4 \times 10^{-5}$ for the magnitude of spatial fluctuations, which are thus typically small.  Conditions more favorable for the observation of these spatial fluctuations can nevertheless be achieved by lowering  the mean free path. For instance, nichrome is a nickel-chromium alloy used in resistive wires. It has $\ell = 4$ nm~\cite{Davidson1976} at room temperature and $n = 9 \times 10^{28} / \text{m}^3$. This leads to spatial fluctuations that are than an order of magnitude larger than for gold. Alternatively, disordered thick films traditionally used in mesoscopic physics could be promising candidates for experimentally unveiling these fluctuations \cite{Lee85}.

\acknowledgments
We thank Christophe Texier for useful discussions on the physics of conductivity correlations in
disordered metals.

\end{document}